\documentclass[12pt]{article}
\usepackage{epsfig}
\usepackage{amssymb}
\usepackage{amsmath}
\usepackage{amsfonts}

\newcommand{\R}{\mathbb{R}}
\newcommand{\C}{\mathbb{C}}
\newcommand{\Z}{\mathbb{Z}}

\newcommand{\fa}{\mathfrak{a}}
\newcommand{\fb}{\mathfrak{b}}
\newcommand{\fc}{\mathfrak{c}}

\newcommand{\fn}{\mathfrak{n}}

\newcommand{\be}{\begin{equation}}
\newcommand{\ee}{\end{equation}}
\newcommand{\bea}{\begin{eqnarray}}
\newcommand{\eea}{\end{eqnarray}}

\newcommand{\kt}{\rangle}
\newcommand{\br}{\langle}

\newcommand{\ed}{\end{document}}

\newcommand{\pbr}{\prec\!}
\newcommand{\pkt}{\!\succ}

\newcommand{\bi}{\begin{itemize}}
\newcommand{\ei}{\end{itemize}}

\begin{document}

\title{Time-Dependent Pseudo-Hermitian
Hamiltonians Defining a Unitary Quantum System and Uniqueness\\ of
the Metric Operator}
\author{\\
Ali Mostafazadeh
\\
\\
Department of Mathematics, Ko\c{c} University,\\
34450 Sariyer, Istanbul, Turkey\\ amostafazadeh@ku.edu.tr}
\date{ }
\maketitle

\begin{abstract}

The quantum measurement axiom dictates that physical observables
and in particular the Hamiltonian must be diagonalizable and have
a real spectrum. For a time-independent Hamiltonian (with a
discrete spectrum) these conditions ensure the existence of a
positive-definite inner product that renders the Hamiltonian
self-adjoint. Unlike for a time-independent Hamiltonian, this does
not imply the unitarity of the Schr\"odinger time-evolution for a
general time-dependent Hamiltonian. We give an additional
necessary and sufficient condition for the unitarity of
time-evolution. In particular, we obtain the general form of a
two-level Hamiltonian that fulfils this condition. We show that
this condition is geometrical in nature and that it implies the
reality of the adiabatic geometric phases. We also address the
problem of the uniqueness of the metric operator.

\vspace{5mm}

\vspace{2mm}


\end{abstract}


\section{Introduction}

In quantum mechanics, the inner product of the Hilbert space of a
quantum system is not an observable quantity. Because all
(separable) Hilbert spaces are unitary equivalent, a convenient
choice is to fix the inner product to be the $L^2$-inner product
$\br\cdot|\cdot\kt$ and formulate the theory on the resulting
Hilbert space ${\cal H}$. In recent years, it has become clear
that one can also formulate a consistent quantum theory that
employs certain non-self-adjoint Hamiltonian operators $H:{\cal
H}\to{\cal H}$, \cite{geyer-scholtz,p2,bender-prl-2002}. The
latter are operators that can be made self-adjoint, if one selects
a new inner product $\br\cdot,\cdot\kt_+$. This in turn implies
that $H$ is a diagonalizable operator with a real
spectrum.\footnote{The diagonalizability of $H$ means that it has
a complete set of eigenvectors.} The physical Hilbert space ${\cal
H}_{\rm phys}$ is obtained by endowing the span of the
eigenvectors of $H$ with the inner product $\br\cdot,\cdot\kt_+$
and completing the resulting inner product space \cite{jpa-2003}.
The observables are identified with self-adjoint operators acting
in ${\cal H}_{\rm phys}$,
\cite{jpa-2003,jpa-2004b}.\footnote{These constructions do not
lead to a generalization of quantum mechanics but to another of
its equivalent representations. This is because one can describe
the very same systems using self-adjoint Hamiltonians and
self-adjoint observables acting in ${\cal H}$, \cite{jpa-2003}.}
The purpose of this paper is to show that the above mentioned
developments do not directly extend to diagonalizable
time-dependent Hamiltonians having a real spectrum.

Time-dependent Hamiltonian operators have many applications in
non-relativistic quantum mechanics. They are also indispensable in
the Hamiltonian formulation of quantum field theories. Therefore a
consistent treatment of pseudo-Hermitian (and in particular ${\cal
PT}$-symmetric \cite{bender-plb}) quantum field theories calls for
a careful study of the extension of the methods of
pseudo-Hermitian quantum mechanics \cite{jpa-2004b} to
time-dependent Hamiltonians. Such Hamiltonians arise naturally in
quantum cosmological applications of pseudo-Hermitian quantum
mechanics \cite{cqg-2003}.

In \cite{fring} the authors consider particular examples of
time-dependent pseudo-Hermitian Hamiltonians that admit a
time-independent (positive-definite) metric operator. In this
article, we will refer to such Hamiltonians as quasi-stationary
and show that a direct extension of the methods of
pseudo-Hermitian quantum mechanics to time-dependent Hamiltonians
is possible provided that they are quasi-stationary. We will then
derive a necessary and sufficient condition under which a given
time-dependent diagonalizable operator with a real and discrete
spectrum is quasi-stationary.

In the remainder of this section we give a brief review of the
spectral methods used in the construction of the metric operators
for a diagonalizable operator with a real and discrete spectrum
\cite{p1,p2}.

Let ${\cal H}$ be a separable (reference) Hilbert space with
($L^2$-) inner product $\br\cdot|\cdot\kt$, and $H:{\cal
H}\to{\cal H}$ be a diagonalizable (Hamiltonian) operator with a
real and discrete spectrum. The diagonalizability of $H$ and the
reality of its spectrum are necessary conditions for the
applicability of the standard quantum measurement theory
\cite{cjp-2006}. The discreteness of the spectrum of $H$ is a
simplifying assumption that could be relaxed depending on the
particular operator in question \cite{cqg-2003,jmp-2005}.

As shown in \cite{p1}, an operator $H$ with the above-mentioned
properties is necessarily pseudo-Hermitian, i.e., there is a
Hermitian invertible (pseudo-metric) operator $\eta:{\cal
H}\to{\cal H}$ satisfying
    \be
    H^\dagger=\eta H\eta^{-1}.
    \label{ph-indef}
    \ee
Furthermore, among the infinity of pseudo-metric operators $\eta$
satisfying this condition there are positive-definite operators
$\eta_+$ that can be used to construct a positive-definite inner
product \cite{p2},\footnote{Equation~(\ref{ph-indef}) was
initially considered by Pauli \cite{pauli} in trying to formalize
an idea due to Dirac \cite{dirac} that later led to the
development of the indefinite-metric quantum theories
\cite{indef-phys}. But Pauli and others who contributed to this
development only considered the case that $\eta$ was a fixed
(given) indefinite operator. The idea of treating (\ref{ph-indef})
as an equation for $\eta$ and realizing that for some $H$ one can
choose a positive-definite operator among all possible $\eta$'s,
that is embraced in \cite{p2}, has its origin in the particular
definition of pseudo-Hermiticity given in \cite{p1}. This is
different from the old notion of ``pseudo-Hermiticity'' used in
indefinite-metric theories. The latter is known as $J$-Hermiticity
in mathematical literature \cite{indef-math}. For a detailed
discussion see \cite{cjp-2006} and reference 1 therein.}
    \be
    \br\cdot,\cdot\kt_+:=\br\cdot|\eta_+\cdot\kt.
    \label{inn}
    \ee
In view of this relation and the $\eta_+$-pseudo-Hermiticity of
$H$, i.e.,
    \be
    H^\dagger=\eta_+ H\eta_+^{-1},
    \label{ph}
    \ee
$H$ is self-adjoint with respect to the inner product
$\br\cdot,\cdot\kt_+$. The operator $\eta_+$ and the corresponding
inner product $\br\cdot,\cdot\kt_+$ that defines the physical
Hilbert space ${\cal H}_{\rm phys}$ of the system are not unique
\cite{npb-2002}.\footnote{The so-called ${\cal CPT}$-inner
products \cite{bender-prl-2002,bender-prd-2004,geyer-znojil} that
can be constructed for typical ${\cal PT}$-symmetric Hamiltonians
form a special class of the inner products $\br\cdot,\cdot\kt_+$,
\cite{jmp-2003}.}

An important observation made in \cite{p2} is that $H$ is related
to a Hermitian operator $h:{\cal H}\to{\cal H}$ via a similarity
transformation. For example, we can choose $\eta_+^{1/2}$ to
perform such a similarity transformation and define $h$ as
    \be
    h:=\eta_+^{1/2}H\eta_+^{-1/2}.
    \label{h}
    \ee
In fact, viewing $H$ and $h$ as acting in ${\cal H}_{\rm phys}$
and ${\cal H}$ respectively and viewing $\eta_+^{1/2}$ as an
operator mapping ${\cal H}_{\rm phys}$ to ${\cal H}$, we find that
indeed $\eta_+^{1/2}$ is a unitary operator\footnote{This means
that for all $\psi,\phi\in {\cal H}_{\rm phys}$,
$\br\eta^{1/2}\psi|\eta^{1/2}\phi\kt=\br\psi,\phi\kt_+$.}, and $h$
and $H$ are unitary equivalent \cite{jpa-2003}. This in turn
allows for the formulation of the theory in terms of the Hermitian
Hamiltonian $h$ within the framework of conventional quantum
mechanics.\footnote{For the cases that ${\cal H}$ is
infinite-dimensional and $H$ has the standard (kinetic+potential)
form, the equivalent Hermitian Hamiltonian $h$ is a typically
nonlocal operator \cite{jpa-2004b,jmp-2005}. But there are
specific cases that it turns out to be local \cite{geyer-cjp}.}

Because $H$ is assumed to be diagonalizable, one can construct a
complete biorthonormal system $\{\psi_n,\phi_n\}$ for the Hilbert
space such that $\psi_n$ and $\phi_n$ are respectively the
eigenvectors of $H$ and $H^\dagger$ with eigenvalue $E_n$,
\cite{p2},
    \be
    H|\psi_n\kt=E_n|\psi_n\kt,~~~~~~
    H^\dagger|\phi_n\kt=E_n|\phi_n\kt.
    \label{eg-va}
    \ee
We also recall that ``biorthonormality'' means
    \be
    \br\phi_n|\psi_m\kt=\delta_{mn},~~~~~~~
    \sum_n|\psi_n\kt\br\phi_n|=1,
    \label{biortho}
    \ee
where we use Dirac's bra-ket notation in ${\cal H}$, $\delta_{mn}$
stands for the Kronecker delta symbol, and $1$ denotes the
identity operator. In terms of the biorthonormal system
$\{\psi_n,\phi_n\}$, we can construct the following metric
operator.
    \be
    \eta_+=\sum_n |\phi_n\kt\br\phi_n|.
    \label{eta=}
    \ee
Indeed every metric operator that satisfies (\ref{ph}) can be
expressed in this form for some biorthonormal system
$\{\psi_n,\phi_n\}$ fulfilling (\ref{eg-va}), \cite{jmp-2003}.

\section{Quasi-Stationary Pseudo-Hermitian Hamiltonians}

Consider a Hamiltonian operator $H[R]$ that is parameterized by
points $R$ of a parameter space $M$. Suppose that $H[R]$ is
diagonalizable and has a real and discrete spectrum for all $R\in
M$. Then there is an $R$-dependent biorthonormal system
$\{\psi_n,\phi_n\}$ satisfying (\ref{eg-va}) with $H=H[R]$ for all
$R\in M$. Furthermore, $H[R]$ is $\eta_+$-pseudo-Hermitian for a
metric operator $\eta_+$ of the form (\ref{eta=}). We will use the
following notation to make the $R$-dependence of $\psi_n,\phi_n$
and $\eta_+$ explicit: $|\psi_n,R\kt:=|\psi_n\kt$,
$|\phi_n,R\kt:=|\phi_n\kt$, and $\eta_+[R]:=\eta_+$.

Next, let $T\in\R^+$, and $\gamma:[0,T]\to M$ be a smooth curve in
$M$ that determines the time-dependence of the parameters $R$ and
the Hamiltonian according to $R(t):=\gamma(t)$ and
$H(t):=H[R(t)]$, respectively. Let us also introduce the
abbreviated notation: $|\psi_n,t\kt:=|\psi_n,R(t)\kt$,
$|\phi_n,t\kt:=|\phi_n,R(t)\kt$, and $\eta_+(t):=\eta_+[R(t)]$.

As pointed out in \cite{cqg-2003}, the
$\eta_+(t)$-pseudo-Hermiticity of $H(t)$ does not generally ensure
the unitarity of the Schr\"odinger time-evolution determined by
$H(t)$ even if we define the Hilbert space using the inner product
$\br\cdot,\cdot\kt_+:=\br\cdot|\eta_+(t)\,\cdot\kt$. To see this,
we denote the time-evolution operator of the system by $U(t)$,
i.e., the operator satisfying the defining relations:
    \be
    i\hbar\,\frac{d}{dt}\,U(t)=H(t)U(t),~~~~~U(0)=1.
    \label{sch-eq}
    \ee
Let $\pbr\cdot,\cdot\pkt$ be a general possibly time-dependent
positive-definite inner product on ${\cal H}$. Then we can always
express $\pbr\cdot,\cdot\pkt$ in terms of a possibly
time-dependent metric operator $\xi_+(t)$ according to \cite{kato}
    \be
    \pbr\cdot,\cdot\pkt=\br\cdot|\xi_+(t)\,\cdot\kt.
    \label{pinn}
    \ee
Suppose that $\psi(t)$ and $\phi(t)$ are arbitrary evolving state
vectors;
    \be
    \psi(t):=U(t)\psi(0),~~~~~~
    \phi(t):=U(t)\phi(0).
    \label{evo}
    \ee
Then the unitarity of time-evolution with respect to the inner
product $\pbr\cdot,\cdot\pkt$ means that $\pbr\psi(t),\phi(t)\pkt$
does not depend on $t$. In view of (\ref{pinn}) and (\ref{evo}),
this condition is equivalent to
    \be
    \xi_+(t)=U(t)^{-1\dagger}\xi_+(0)U(t)^{-1}.
    \label{e1}
    \ee
Differentiating both sides of this equation and using
(\ref{sch-eq}) we find
    \be
    H(t)^\dagger=\xi_+(t)H(t)\xi_+(t)^{-1}-i\xi(t)\dot\xi(t)^{-1},
    \label{ph-add}
    \ee
where a dot denotes a time-derivative. Equation~(\ref{ph-add})
shows that $H(t)$ is $\xi_+$-pseudo-Hermitian if and only if
$\xi_+$ is time-independent.

The requirement of the unitarity of time-evolution demands that
the inner product of the Hilbert space be defined by a metric
operator fulfilling (\ref{ph-add}). On the other hand quantum
measurement theory (projection axiom) requires $H(t)$ to be
self-adjoint with respect to the defining inner product of the
physical Hilbert space of the system. These two constraints imply
that \emph{a time-dependent Hamiltonian operator $H(t)$ defines a
consistent unitary quantum system if and only if $H(t)$ is
$\eta_+$-pseudo-Hermitian for a time-independent metric operator
$\eta_+$}. We will call such a Hamiltonian
\emph{quasi-stationary}.

Requiring $H(t)$ to be quasi-stationary puts a sever restriction
on its eigenvectors. To see this we choose an arbitrary metric
operator $\eta_+$ satisfying (\ref{ph}), use an appropriate
biorthonormal system $\{|\psi_n,t\kt,|\phi_n,t\kt\}$ to express it
in the form (\ref{eta=}), and demand that the time-derivative of
both sides of this equation vanishes. In view of (\ref{biortho}),
this yields
    \be
    {\cal A}_{mn}(t)={\cal A}_{nm}(t)^*,
    \label{AA}
    \ee
where
    \be
    {\cal A}_{mn}(t):=i\br\phi_m,R|\frac{d}{dt}|\psi_n,R\kt.
    \label{A=}
    \ee
If we assume that all the parameters $R$ describe physical
situations, the condition (\ref{AA}) must be met for all possible
curves $\gamma:[0,T]\to M$. This is equivalent to
    \be
    A_{mn}[R]=A_{nm}[R]^*,
    \label{AA2}
    \ee
where
    \be
    A_{mn}[R]:=i\br\phi_m,R|d|\psi_n,R\kt:=\sum_{i=1}^n
    i\br\phi_m,R|\frac{\partial}{\partial R_i}|\psi_n,R\kt\,dR_i.
    \label{A=R}
    \ee
The one-form $A_{nn}[R]$ is the Berry's connection one-form for a
diagonalizable non-Hermitian Hamiltonian \cite{garrison-wright}.
Therefore, a simple implication of (\ref{AA2}) is that the
adiabatic geometric phase angles for the system are
real.\footnote{The dynamical phase angles are also real, because
$H(t)$ has a real spectrum.} This is actually to be expected,
because the system admits a Hermitian representation in terms of a
Hermitian Hamiltonian, and like other physical quantities the
geometric phase angles can be calculated in the Hermitian
representation where they are clearly real.

It is important to observe that the condition that $H(t)$ be
quasi-stationary is equivalent to the requirement of the existence
of a biorthonormal system $\{|\psi_n,R\kt,|\phi_n,R\kt\}$ such
that $|\psi_n,R\kt$ and $|\phi_n,R\kt$ are respectively the
eigenvectors of $H[R]$ and $H[R]^\dagger$ and that (\ref{AA2}) is
satisfied. Note also that this condition is not sensitive to the
duration of the evolution of the system and is completely
geometrical in nature.

\section{Two-Level Quasi-Stationary Pseudo-Hermitian Hamiltonians}

Consider the case that ${\cal H}$ is the two-dimensional complex
vector space $\C^2$ endowed with the Euclidean ($L^2$-) inner
product: $\br\psi|\phi\kt:=\sum_{a=1}^2\psi^{a*}\phi^a$, where
$\psi:=\mbox{\scriptsize$\left(\begin{array}{c}\psi^1\\
\psi^2\end{array}\right)$}$, $\phi:=\mbox{\scriptsize$
\left(\begin{array}{c}\phi^1\\
\phi^2\end{array}\right)$}$, and $\psi^a,\phi^a\in\C$ for all
$a\in \{1,2\}$. In the standard basis $\mbox{\scriptsize$
\left\{\left(\begin{array}{c}1\\
0\end{array}\right),\left(\begin{array}{c}0\\
1\end{array}\right)\right\}$}$ of $\C^2$, we can represent the
most general diagonalizable Hamiltonian operator $H$ with a real
spectrum as
    \be
    H=q\,I+H_0,~~~~~~~~~~~~
    H_0=\left(\begin{array}{cc}\fa & \fb\\ \fc & -\fa
    \end{array}\right),\label{H=}
    \ee
where $q\in\R$, $I$ is the $2\times 2$ unit matrix,
$\fa,\fb,\fc\in\C$, and $\fa^2+\fb\fc$ is real and nonnegative,
\cite{npb-2002,tjp-2006}. The problem of finding the most general
metric operator $\eta_+$ for such a Hamiltonian has been
completely solved in \cite{tjp-2006}.\footnote{The two-level
pseudo-Hermitian Hamiltonians have also been considered in
\cite{jpa-2003,finite,geyer-znojil}.}

Reparameterizing $H_0$ in the form \cite{kvitsimsky-putterman}
    \be
    H_0=E\left(\begin{array}{cc}\cos\theta & e^{-i\varphi}
    \sin\theta\\ e^{i\varphi}
    \sin\theta & -\cos\theta\end{array}\right),
    \label{H-zero}
    \ee
where $E:=\sqrt{\fa^2+\fb\fc}\in [0,\infty)$,
$\theta,\varphi\in\C$, $\Re(\theta)\in[0,\pi]$,\footnote{$\Re$ and
$\Im$ stand for the real and imaginary part of their argument
respectively.} and $\Re(\varphi)\in[0,2\pi)$, we can express the
most general biorthonormal system associated with $H$ as
    \bea
    |\psi_1 \kt=\fn_1^{-1*}
    \left(\begin{array}{c}\cos\mbox{$\frac{\theta}{2}$}\\
    e^{i\varphi}\sin\mbox{$\frac{\theta}{2}$}\end{array}\right),~~~~
    |\psi_2 \kt=\fn_2^{-1*}
    \left(\begin{array}{c}\sin\mbox{$\frac{\theta}{2}$}\\
    -e^{i\varphi}\cos\mbox{$\frac{\theta}{2}$}\end{array}\right),
    \label{psi-}\\
    |\phi_1 \kt=\fn_1
    \left(\begin{array}{c}\cos\mbox{$\frac{\theta^*}{2}$}\\
    e^{i\varphi^*}\sin\mbox{$\frac{\theta^*}{2}$}\end{array}\right),~~~~
    |\phi_2 \kt=\fn_2
    \left(\begin{array}{c}\sin\mbox{$\frac{\theta^*}{2}$}\\
    -e^{i\varphi^*}\cos\mbox{$\frac{\theta^*}{2}$}\end{array}\right),
    \label{phi-}
    \eea
where $\fn_1,\fn_2\in\C-\{0\}$ are arbitrary. The eigenvalues of
$H$ (and $H^\dagger$) are given by $E_1=q+E$ and $E_2=q-E$.
Substituting (\ref{phi-}) in (\ref{eta=}) and carrying out the
necessary calculations, we find the following expression for the
most general metric operator $\eta_+$ such that $H$ is
$\eta_+$-pseudo-Hermitian \cite{tjp-2006}.
    \be
    \eta_+=k\left(\begin{array}{cc}au+b & \lambda^*\\
    \lambda & e^{2\Im(\varphi)}(a+bu)\end{array}\right),
    \label{eta=2}
    \ee
where $k:=|\fn_2|^2$ and $u:=|\fn_1/\fn_2|^2$ are arbitrary
positive real parameters manifesting the non-uniqueness of
$\eta_+$, and
    \[a:=|\cos\mbox{$\frac{\theta}{2}$}|^2,~~~
    b:=|\sin\mbox{$\frac{\theta}{2}$}|^2,~~~
    \lambda:=e^{i\varphi}(u\zeta^*-\zeta),~~~
    \zeta:=
    \sin\mbox{$\frac{\theta}{2}$}\,\cos\mbox{$\frac{\theta^*}{2}$}.\]

Clearly the Hamiltonian $H$ depends on the six real parameters
$q,E,\Re(\theta),\Im(\theta),\Re(\varphi)$, and $\Im(\varphi)$
that can be collectively denoted by $R$. As $q$ and $E$ do not
enter the expression for the biorthonormal system, the condition
that $H$ be quasi-stationary only restricts $\theta$ and
$\varphi$. Inserting (\ref{psi-}) and (\ref{phi-}) in (\ref{AA})
and simplifying the resulting equations, we can express this
condition in the form of the following system of ordinary
differential equations.
    \bea
    \Im[\,\sin^2(\mbox{$\frac{\theta^*}{2}$})\:\dot\varphi]+
    \dot\nu_1&=&0,
    \label{eq1}\\
    \Im[\,\cos^2(\mbox{$\frac{\theta^*}{2}$})\:\dot\varphi]+
    \dot\nu_2&=&0,
    \label{eq2}\\
    \Im(\dot\theta)-\mu\,\Re[\,\sin(\theta)\,\dot\varphi]&=&0,
    \label{eq3}\\
    \mu\,\Re(\dot\theta)+\Im[\,\sin(\theta)\,\dot\varphi]&=&0,
    \label{eq4}
    \eea
where $\nu_a:=\ln|\fn_a|$ for $a\in\{1,2\}$ and $\mu:=
\mbox{\small$\frac{|\fn_1|^2-|\fn_2|^2}{|\fn_1|^2+|\fn_2|^2}$}$.
Equations~(\ref{eq1}) and (\ref{eq2}) may be solved to express
$\fn_1$ and $\fn_2$ in terms of $\theta$ and $\varphi$.
Substituting the result in Equations~(\ref{eq3}) and (\ref{eq4})
yields two real equations for the four unknown functions
$\Re(\theta),\Im(\theta),\Re(\varphi)$ and $\Im(\varphi)$. Note
that equations (\ref{eq1}) -- (\ref{eq4}) are
time-reparameterization-invariant; we can eliminate $t$ from these
equations and express them in terms of any of the real parameters
of the system, e.g., $\Re(\varphi)$.

By construction, solving (\ref{eq1}) -- (\ref{eq4}) is equivalent
to demanding that $\eta_+$ as given by (\ref{eta=2}) is a
constant. This means that both $H(0)$ and $H(t)$ (for any
$t\in[0,T]$) are $\eta_+$-pseudo-Hermitian. Therefore, we can
obtain a characterization of quasi-stationary Hamiltonians $H(t)$
by setting $t=0$ in (\ref{eta=2}) and finding the form of $H(t)$
that is $\eta_+$-pseudo-Hermitian. This allows for an algebraic
solution of the system of equations (\ref{eq1}) -- (\ref{eq4}).

The problem of finding the general form of an
$\eta_+$-pseudo-Hermitian operator for metric operators of the
form (\ref{eta=2}) has also been solved in \cite{tjp-2006}. Here
we summarize the result. There are two possibilities:
    \begin{enumerate}
    \item $\lambda(0)=0$ (i.e., $\eta_+$ is diagonal), which
    corresponds to the cases:
    \textbf{(1.a)} $\theta(0)=0$; \textbf{(1.b)} $\theta(0)\in\R$ and
    $u =1$. For theses cases, we have
        \be
        \Im[\,\fa(t)]=0,~~~~~~~~~~~
        \fc(t)=\left[\frac{a(0)u +b(0)}{a(0)+b(0)u }\right]\,
        e^{-2\Im[\varphi(0)]}\,\fb(t)^*.
        \label{sol1}
        \ee
    Here $q(t)$, $\Re[\fa(t)]$, $\Re[\fb(t)]$ and
    $\Im[\fb(t)]$ are arbitrary real-valued functions.
    \item $\lambda(0)\neq 0$ (i.e., $\eta_+$ is not diagonal),
    which corresponds to the cases:
    \textbf{(2.a)} $u \neq 1$ and $\theta(0)\neq 0$;
    \textbf{(2.b)} $\theta(0)\notin\R$.
    For theses cases, we have
        \bea
        \fb(t)&=&\lambda(0)^{-1}\left\{f(t)+ir\,\Im[\fa(t)]\right\},
        \label{sol2-1}\\
        \fc(t)&=&[r\lambda(0)^*]^{-1}\left\{s\,f(t)-2|\lambda(0)|^2\,
        \Re[\fa(t)]-irs\,\Im[\fa(t)]\right\},
        \label{sol2-2}
        \eea
    where $r:=e^{2\Im[\varphi(0)]}[a(0)+b(0)u $], $s:=a(0)u +
    b(0)$, and $f$ is an arbitrary real-valued function. Again,
    $H$ has four functional real degrees of freedom, namely
    $q(t)$, $f(t)$, $\Re[\fa(t)]$ and $\Im[\fa(t)]$.
    \end{enumerate}
Note that in both cases $u $ is a positive real constant that can
sometimes be determined by setting $t=0$ in (\ref{sol1}) --
(\ref{sol2-2}). If this fixes $u $, the metric operator $\eta_+$
is uniquely determined up to the unimportant multiplicative
constant $k $. Otherwise, similarly to the case of a
time-independent Hamiltonian, the determination of $\eta_+$
amounts to making a choice for $u $, \cite{tjp-2006}. We will
return to this problem in Section~4.

In practice we can employ the above results as follows. Given a
(non-diagonal) time-dependent $2\times 2$ matrix Hamiltonian
$H(t)$, we determine whether it is diagonalizable and has a real
spectrum by examining its trace and the determinant of its
traceless part \cite{npb-2002}. If both of these quantities are
real and the latter is negative, $H(t)$ is diagonalizable and has
a real spectrum.\footnote{In this case $H(t)$ can be put in the
form (\ref{H=}).} But as we explained in the preceding section,
this is not sufficient for formulating a consistent quantum theory
using $H(t)$. In addition, the Hamiltonian must be
quasi-stationary. To see if this is the case we examine its
diagonal entries. If they are both real, then the Hamiltonian is
quasi-stationary if (\ref{sol1}) holds with $u =1$. If at least
one of the diagonal entries is not real, then the Hamiltonian is
quasi-stationary provided that it satisfies (\ref{sol2-1}) and
(\ref{sol2-2}).

\section{Uniqueness of the Metric Operator}

For a time-independent diagonalizable Hamiltonian $H$ with a real
spectrum, the metric operator $\eta_+$ that makes $H$,
$\eta_+$-pseudo-Hermitian is not unique \cite{npb-2002,jmp-2003}.
In general one must fix a metric operator $\eta_+$ directly
\cite{jpa-2004b} or alternatively select a set of so-called
compatible irreducible operators and demand that all of these
operators be $\eta_+$-pseudo-Hermitian
\cite{geyer-scholtz,tjp-2006}. The latter will fix $\eta_+$ up to
an unimportant multiplicative positive real constant.

The situation is different for a time-dependent Hamiltonian. The
requirement that a generic time-dependent Hamiltonian $H(t)$ be
quasi-stationary, i.e., $H(t)$ be $\eta_+$-pseudo-Hermitian for a
constant $\eta_+$, will fix $\eta_+$ (again up to an unimportant
multiplicative positive constant). To see this, consider a general
quasi-stationary Hamiltonian $H(t)$ and suppose that $\eta_+$ is a
constant metric operator such that $H(t)$ is
$\eta_+$-pseudo-Hermitian. Then, for all $t\in[0,T]$,
    \be
    H(t)^\dagger=\eta_+H(t)\eta_+^{-1}.
    \label{z2}
    \ee
Setting $t=0$ in this relation implies that $H(0)$ is
$\eta_+$-pseudo-Hermitian. Differentiating both sides of
(\ref{z2}) successively, setting $t=0$, and defining $O_0=H(0)$
and for all $n\in\Z^+$, $O_n:=\frac{d^n}{dt^n} H(t)\big|_{t=0}$,
we find that $(O_0,O_1,O_2,\cdots)$ is an infinite sequence of
$\eta_+$-pseudo-Hermitian operators. Assuming that $O_n$'s do not
share a common eigenvector, which is true for a generic
Hamiltonian $H(t)$, the sequence $(O_0,O_1,O_2,\cdots)$ includes
among its terms an irreducible set of operator. Therefore,
according to the uniqueness theorem proven in
\cite{geyer-scholtz}, $\eta_+$ is unique up to a constant factor.
For the two-dimensional systems considered in Section~3, this
manifests itself through the fact that for a generic Hamiltonian
the parameter $u $ that represents the arbitrariness in the choice
of $\eta_+$ is fixed by setting $t=0$ in Equations (\ref{sol1}) --
(\ref{sol2-2}).

To see how this is done, consider the case that $q=0$ and
$\theta=\mbox{$\frac{\pi}{2}$}$, i.e.,
    \be
    H(t)=\left(\begin{array}{cc}
    0 & \fb(t)\\
    \fc(t) & 0\end{array}\right).
    \label{H-special}
    \ee
Then the condition that $H(t)$ is a (nonzero) diagonalizable
operator with a real spectrum takes the form $\fb\fc\in\R^+$. To
obtain the form of $\fb$ and $\fc$ for which $H(t)$ is
quasi-stationary, we consider the following two possibilities.
    \begin{itemize}
    \item[] (i) $u =1$: In this case $\lambda(0)= 0$ and we should
    enforce (\ref{sol1}). But we can easily check that this does not
    put any restriction on $\fb$ and $\fc$.
    \item[] (ii) $u \neq 1$: In this case $\lambda(0)\neq 0$ and
    we should enforce (\ref{sol2-1}) and (\ref{sol2-2}). They give
        \be
        \fb=\frac{2 e^{-i\varphi(0)}f(t)}{u_0-1},~~~~
        \fc=\frac{2 e^{i\varphi(0)}f(t)}{u_0-1},
        \label{ww1}
        \ee
    where $u $ is an arbitrary positive real number different
    from 1, $\varphi(0)$ is an arbitrary complex number, and
    $f(t)$ is an arbitrary real-valued function. Note that
    according to (\ref{ww1}),
        \be
        H(t)=f(0)^{-1}f(t)\,H(0).
        \label{trivial}
        \ee
    This is the trivial case, where the eigenvectors of $H(t)$
    happen to be time-independent.
    \end{itemize}
The above analysis shows that if a Hamiltonian of the form
(\ref{H-special}) does not satisfy (\ref{trivial}) for any
real-valued function $f$, then we must choose $u =1$. This in turn
means that the metric operator $\eta_+$ is determined uniquely up
to the constant factor $k $. But if we can satisfy (\ref{trivial})
for some $f$, then $u $ may be chosen arbitrarily. In the latter
case, similarly to the case of a time-independent Hamiltonian in
order to fix $\eta_+$, we must also make a choice for $u $.

\section{Concluding Remarks}

In this paper we have shown that unlike for a time-independent
Hamiltonian operator, the conditions of diagonalizability and
reality of the spectrum of a time-dependent Hamiltonian do not
generally guarantee the unitarity of the corresponding
time-evolution. The latter puts a further restriction on the
choice of the Hamiltonian. We have examined the general form of
this restriction, elucidated its geometric nature, and given a
complete characterization of time-dependent $2\times 2$ matrix
Hamiltonians that define consistent quantum theories. We have also
shown that, in contrast to the case of a time-independent
Hamiltonian, a generic time-dependent Hamiltonian that is capable
of defining a consistent quantum theory determines the metric
operator and the inner product of the physical Hilbert space
uniquely (up to a physically irrelevant multiplicative numerical
factor).



\ed
\begin{thebibliography}{99}
\bibitem{geyer-scholtz} F.~G.~Scholtz, H.~B.~Geyer, and
F.~J.~W.~Hahne, Ann.\ Phys.\ (NY) {\bf 213} 74 (1992).
\bibitem{p2} A.~Mostafazadeh, J.\ Math.\ Phys.\ {\bf 43}, 2814 and
3944 (2002).
\bibitem{bender-prl-2002} C.~M.~Bender, D.~C.~Brody and H.~F.~Jones,
Phys.\ Rev.\ Lett.\ {\bf 89}, 270401 (2002).
\bibitem{jpa-2003} A.~Mostafazadeh, J.~Phys.~A {\bf 36}, 7081
(2003); and Czech J.~Phys.\ {\bf 54}, 1125 (2004).
\bibitem{jpa-2004b} A.~Mostafazadeh and A.\ Batal, J.~Phys.~A {\bf 37},
11645 (2004); A.~Mostafazadeh, J.~Phys.~A {\bf 38}, 3213 (2005).
\bibitem{bender-plb} C.~M.~Bender, I.~Cavero-Pelaez,
K.~A.~Milton, and K.~V.~Shajesh, Phys.\ Lett.~B {\bf 613}, 97
(2005); C.~M.~Bender, H.~F.~Jones, and R.~J.~Rivers, Phys.\
Lett.~B {\bf 625}, 333 (2005); and references therein.
\bibitem{cqg-2003} A.~Mostafazadeh, Class.\ Quantum Grav.\ {\bf 20},
155 (2003); and Ann.~Phys.~(N.Y.) {\bf 309}, 1 (2004).
\bibitem{fring} C.~Figueira de Morisson and A.~Fring, J.~Phys.~A {\bf 39},
9269 (2006).
\bibitem{p1} A.~Mostafazadeh, J.\ Math.\ Phys.\ {\bf 43}, 205
(2002).
\bibitem{cjp-2006} A.~Mostafazadeh, Czech J.~Phys.\ {\bf 56},
919 (2006).
\bibitem{jmp-2005} A.~Mostafazadeh, J.~Math.~Phys.\ {\bf 46}, 102108
(2005); and J.~Phys.~A {\bf 39}, 13495 (2006).
\bibitem{pauli} W.~Pauli, Rev.\  Mod.\ Phys., {\bf 15}, 175
(1943).
\bibitem{dirac} P.\ A.\ M.\ Dirac, Proc.\ Roy.\ Soc.\ London A
{\bf 180}, 1 (1942).
\bibitem{indef-phys} K.~L.~Nagy, {\em State Vector Spaces with
Indefinite Metric Quantum Field Theory}, Noordhoff, Groningen,
Netherlands, 1966; N.~Nakanishi, Suppl.\ Prog.\ Theor.\ Phys.\
{\bf 51}, 1 (1972).
\bibitem{indef-math} T.\ Ya.\ Azizov and I.\ S.\ Iokhvidov,
{\em Linear Operators in Spaces with Indefinite Metric} (Wiley,
Chichester, 1989).
\bibitem{npb-2002} A.~Mostafazadeh, Nucl.\ Phys.\ B, {\bf 640}, 419
(2002).
\bibitem{bender-prd-2004} C.~M.~Bender, D.~C.~Brody and H.~F.~Jones,
Phys.\ Rev.\ D {\bf 70}, 025001 (2004).
\bibitem{geyer-znojil} M.~Znojil and H.~B.~Geyer, Phys.\ Lett.~B
{\bf 640}, 52 (2006).
\bibitem{jmp-2003} A.~Mostafazadeh, J.\ Math.\ Phys.\ {\bf 44}, 974
(2003); and J.~Phys.~A {\bf 39}, 10171 (2006).
\bibitem{geyer-cjp} H.~B.~Geyer, F.~G.~Scholtz, I.~Snyman, Czech
J.~Phys.\ {\bf 54}, 1069 (2004); H.~F.~Jones and J.~Mateo,
Phys.~Rev.~D {\bf 73} 085002 (2006); D.~P.~Musumbu, H.~B.~Geyer,
and W.~D.~Heiss, J.~Phys.~A {\bf 40}, F75 (2007).
\bibitem{kato} T.~Kato, {\em Perturbation Theory for Linear
Operators}, Springer, Berlin, 1995.
\bibitem{garrison-wright} J.~C.~Garrison and E.~M.~Wright, Phys.\
Lett.~A {\bf 128}, 177 (1988).
\bibitem{tjp-2006} A.~Mostafazadeh and S.~\"Oz\c{c}elik, Turk.\
J.~Phys.\ {\bf 30}, 437 (2006).
\bibitem{finite} C.~M.~Bender, P.~N.~Meisinger, and Q.~Wang,
J.~Phys.~A {\bf 36}, 6791 (2003); A.~Mostafazadeh, J.\ Math.\
Phys., {\bf 45}, 932 (2004); Y.~Ben-Aryeh, A.~Mann, and I.~Yaakov,
J.~Phys.~A {\bf 37}, 12059 (2004); A.~D.~Dutra, M.~B.~Hott, and
V.~G.~C.~S.~dos Santos, Europhys.\ Lett.~{\bf 71}, 166 (2005);
P.~K.~Ghosh, J.~Phys.~A {\bf 38}, 7313 (2005).
\bibitem{kvitsimsky-putterman} A.~Kvitsimsky and S.~Putterman,
J.\ Math.\ Phys.\ {\bf 32}, 1403 (1991).
\end{thebibliography}
